\documentstyle[12pt]{article}
\begin{document}
\title{\bf $h$--analogue of Newton's binomial formula } 
\author{ H. B. Benaoum~\footnote{ Present and permanent adress : 
Institut f\"ur Physik ,
Johannes Gutenberg--Universit\"at, 55099 Mainz, Germany,~~~
email : benaoum@thep.physik.uni-mainz.de } \\
\small{Institut f\"ur Theoretische Physik, Technische Universit\"at Clausthal} \\
\small{Leibnizstrasse 10, 38678 Clausthal--Zellerfeld, Germany} \\
}
\date{14 July 1998 \\
Revised 06 October 1998}
\maketitle
~\\
\abstract{In this letter, the $h$--analogue of 
Newton's binomial formula is obtained 
in the $h$--deformed quantum plane which does not have any $q$--analogue. 
For $h=0$, this is just the usual one as it should be. Furthermore, the 
binomial coefficients reduce to $\frac{n!}{(n-k)!}$ for $h=1$. \\
Some properties of the $h$--binomial coefficients are also given. \\
Finally, I hope that such results will contribute to an introduction of 
the $h$--analogue of the well--known functions, $h$--special functions 
and $h$--deformed analysis.}
~\\
~\\
~\\
{\bf Published in J. Phys. A : Volume 31 Issue 46 ( 20 November 1998) p. L751
}  \\
~\\
~\\
~\\
MZ--TH/98--41
\newpage
~\\
The study of $q$--analysis appeared in the literature 
very long time ago~\cite{gas}. 
In particular, a $q$--analogue of the Newton's formula, well--known functions 
like $q$--exponential, $q$--logarithm,$\cdots$ etc, 
and the special functions arena's~\cite{gas,nik,koo} 
have been introduced and studied intensively. \\
Such $q$--analogue of these was obtained by taking $q$--commuting variables 
$x,y$ satisfying the relation $x y = q y x$, i.e. $(x, y)$ belongs to the 
Manin plane. \\
In this letter, I will take another direction by introducing the analogue of  
Newton's formula in the $h$--deformed quantum plane~\cite{mad,ben} 
(i.e. $h$--Newton binomial formula ). As far as I know, such a $h$--analogue   
does not exist in the litterature till now and the result will permit in the 
future the introduction of the $h$--analogue of well--known functions, $h$--
special functions and $h$--deformed analysis. \\ 

~\\
Newton's binomial formula is defined as follows :
\begin{eqnarray}
( x + y )^n & = & \sum^n_{k=0} \left( \begin{array}{c}
n \\
k \end{array} \right) y^k x^{n-k}
\end{eqnarray}
where $\left( \begin{array}{c} 
n \\
k \end{array} \right) = \frac{n!}{k! (n-k)!}$  
and it is understood here that the 
coordinate variables $x$ and $y$ commute, i.e. $x y = y x$. \\
A $q$--analogue of (1) for the $q$--commuting coordinates $x$ and $y$ 
satisfying $x y = q y x$ was first stated by Rothe, although its special 
cases were known to L. Euler, see~\cite{bie}, found again by 
Sch\"utzenberger~\cite{sch} long time ago and has been rediscovered       
many times subsequently~\cite{cig}. \\ 
A $q$--analogue of (1) becomes :
\begin{eqnarray}
( x + y )^n & = & \sum^n_{k=0} \left[ \begin{array}{c}
 n \\
 k  \end{array} \right]_q y^k x^{n-k}
\end{eqnarray}
where the $q$--binomial coefficient is given by :
\begin{eqnarray*}
\left[ \begin{array}{c} 
n \\
k \end{array} \right]_q & = & \frac{ (q;q)_n}{ (q;q)_k (q;q)_{n-k}}
\end{eqnarray*}
with 
\begin{eqnarray*}
( a; q )_k & = & (1 - a ) (1 - q a ) \cdots (1 - q^{k-1} a ),~~~
a \in {\bf C},  k \in {\bf N} 
\end{eqnarray*}
~\\
Now consider Manin's $q$--plane $x' y' = q y' x'$.By the following linear 
transformation~( see \cite{mad} and references therein ) :
\begin{eqnarray*}
\left( \begin{array}{c} 
x' \\
y' \end{array} \right) & = & \left( \begin{array}{cc} 
1 & \frac{h}{q-1} \\
0 & 1 \end{array} \right) \left( \begin{array}{c} 
x \\
y \end{array} \right)
\end{eqnarray*}
Manin's $q$--plane changes to $x y - q y x = h y^2$ which for $q=1$ gives 
the $h$--deformed plane :
\begin{eqnarray}
x y & = & y x + h y^2 
\end{eqnarray}
Even though the linear transformation is singular for $q=1$, the resulting 
quantum plane is well-defined. \\
~\\
{\bf \underline{Proposition~1 :} } \\
Let $x$ and $y$ be coordinate variables satisfying (3), then the following 
identities are true :
\begin{eqnarray}
x^k y & = & \sum^k_{r=0} \frac{k!}{(k-r)!} h^r y^{r+1} x^{k-r} \nonumber \\
x y^k & = & y^k x + k h y^{k+1}
\end{eqnarray}
These identities are easily proved by successive use of (3). \\
~\\
{\bf \underline{Proposition~2 :}} ( $h$--binomial formula ) \\
Let $x$ and $y$ be coordinate variables satisfying (3), then we have :
\begin{eqnarray}
( x + y )^n & = & \sum^n_{k=0}  \left[ \begin{array}{c} 
n \\
k \end{array} \right]_h y^k x^{n-k}
\end{eqnarray}
where $\left[ \begin{array}{c} 
n \\
k \end{array} \right]_h$ are the $h$--binomial coefficients given as follows :
\begin{eqnarray}
\left[ \begin{array}{c} 
n \\
k \end{array} \right]_h & = &  
\left( \begin{array}{c}
n \\
k \end{array} \right) 
h^k (h^{-1})_k .
\end{eqnarray}
with $\left[ \begin{array}{c}
n \\
0 \end{array} \right]_h = 1$ and $(a)_k = \Gamma(a + k)/\Gamma(a)$ is the 
shifted factorial. \\
~\\
Proof : \\
We will prove this proposition by recurrence. 
Indeed for $n=1,2$, it is verified. \\
Suppose now that the formula is true for $n-1$, which means :
\begin{eqnarray*}
( x + y ) ^{n-1} & = & \sum^{n-1}_{k=0} \left[ \begin{array}{c} 
n - 1 \\
k \end{array} \right]_h y^k x^{n-1-k},
\end{eqnarray*}
with $\left[ \begin{array}{c} 
n-1 \\
0 \end{array} \right]_h = 1$. \\
To show this for $n$, let first consider the following expansion :
\begin{eqnarray*}
( x + y )^n & = & \sum^n_{k=0} C_{n,k} y ^k x^{n-k} 
\end{eqnarray*}
where $C_{n,k}$ are coefficients depending on $h$. \\
Then, we have : 
\begin{eqnarray*}
( x + y )^n & = & ( x + y ) ( x + y )^{n-1} \nonumber \\
& = & ( x + y ) \sum^{n-1}_{k=0} \left[ \begin{array}{c} 
n -1 \\
k \end{array} \right]_h y^k x^{n-1-k} \nonumber \\
& = & \sum^{n-1}_{k=0} \left[ \begin{array}{c} 
n -1 \\
k \end{array} \right]_h x y^k x^{n-1-k}~+~ \sum^{n-1}_{k=0} \left[ 
\begin{array}{c} 
n -1 \\
k \end{array} \right]_h y^{k+1} x^{n-1-k}.
\end{eqnarray*}
Using the result of the first proposition, we obtain :
\begin{eqnarray*}
( x + y )^n & = & \sum^{n-1}_{k=0} \left[ \begin{array}{c}
n - 1 \\
k \end{array} \right]_h y^k x^{n-k}~+~\sum^{n-1}_{k=0} \left[ \begin{array}{c} 
n - 1 \\
k \end{array} \right]_h ( 1 + k h ) y^{k+1} x^{n-1-k} \nonumber \\
& = & \sum^{n-1}_{k=0} \left[ \begin{array}{c} 
n -1 \\
k \end{array} \right]_h y^k x^{n-k}~+~\sum^n_{k=1} \left[ \begin{array}{c} 
n -1 \\
k -1 \end{array} \right]_h ( 1 + (k-1) h ) y^k x^{n-k}.
\end{eqnarray*}
which yields respectively :
\begin{eqnarray*}
C_{n,0} & = & \left[ \begin{array}{c} 
n - 1 \\
0 \end{array} \right]_h~=~1, \nonumber \\
C_{n,k} & = & \left[ \begin{array}{c} 
n - 1 \\
k \end{array} \right]_h~+~( 1 + (k-1) h ) \left[ \begin{array}{c} 
n - 1 \\
k - 1 \end{array} \right]_h~=~ \left[ \begin{array}{c} 
n \\
k \end{array} \right]_h ,  \nonumber \\
C_{n,n} & = & \left[ \begin{array}{c} 
n - 1 \\
n - 1 \end{array} \right]_h ( 1 + (n -1) h )~=~\left[ \begin{array}{c} 
n \\
n \end{array} \right]_h .
\end{eqnarray*}
This completes the Proof. \\  
~\\
Moreover, the $h$--binomial coefficients obey to the following properties :
\begin{eqnarray}
\left[ \begin{array}{c} 
n \\
k \end{array} \right]_h~ + ~ ( 1 + (k-1) h ) \left[ \begin{array}{c} 
n \\
k - 1 \end{array} \right]_h & = & \left[ \begin{array}{c}
n + 1 \\
k \end{array} \right]_h.
\end{eqnarray}
and 
\begin{eqnarray}
\left[ \begin{array}{c} 
n + 1 \\
k + 1 \end{array} \right]_h & = & \frac{n+1}{k+1} ( 1 + k h ) \left[ 
\begin{array}{c}
n \\
k \end{array} \right]_h.
\end{eqnarray}
In fact, these properties follow from the well--known relations of the 
classical binomial coefficients : 
\begin{eqnarray*}
\left( \begin{array}{c}
n + 1 \\
k \end{array} \right) & = & \left( \begin{array}{c} 
n \\
k \end{array} \right) + \left( \begin{array}{c}
n \\
k - 1 \end{array} \right)
\end{eqnarray*}
and 
\begin{eqnarray*}  
\left( \begin{array}{c} 
n + 1 \\
k \end{array} \right) & = & \frac{n + 1}{k}~\left( \begin{array}{c}
n \\
k - 1 \end{array} \right)
\end{eqnarray*}
upon using $(a)_k = (a + k -1) (a)_{k-1}$, which means that (7) and (8) are  
just a consequence of the known properties of the classical coefficients  
and the shifted factorial. \\
~\\
Now, we make the following remarks. First, for $h=0$ the Newton's binomial 
formula is just the usual one for commuting variables $x y = y x$   
as it should be. \\ 
Second, for $h=1$ the $h=1$--binomial coefficients are :
\begin{eqnarray}
\left[ \begin{array}{c} 
n \\
k \end{array} \right]_{h=1} & = & \frac{n!}{(n-k)!}
\end{eqnarray}
and therefore the $h=1$--analogue Newton's binomial formula becomes :
\begin{eqnarray}
( x + y )^n & = & \sum^n_{k=0} \frac{n!}{(n-k)!} y^k x^{n-k}
\end{eqnarray}
provided that $x y = y x + y^2$. \\
To conclude, we see that the $h$--analogue of Newton's formula in the $h$--
deformed plane has no $q$--analogue. It seems from the structures of the 
$h$--binomial coefficients that the $h$--deformed plane is somewhat "more 
classical" than the $q$--deformed plane.
\section*{acknowlegment}
I'd like to thank the DAAD for its financial support and the referee for his 
remarks. Special thanks go also to Prof. H.-D. Doebner, Dr.  
R. H\"au{\ss}ling for reading the 
manuscript and Prof. F. Scheck for encouragements.
~\\

\end{document}